\documentclass[aps,prl,twocolumn,showpacs,amsmath,superscriptaddress,floatfix]{revtex4}

\usepackage{amsfonts}
\usepackage{amsmath}
\usepackage{amssymb}
\usepackage{graphicx}
\usepackage{color}
\usepackage{verbatim}
\usepackage{lipsum}
\usepackage{dsfont}
\usepackage{hyperref}
\usepackage{upgreek}

\begin{document}
\title{Deterministic single ion implantation of rare-earth ions for nanometer resolution colour center generation}

\author{Karin Groot-Berning}
\email{karin.groot-berning@uni-mainz.de}
\affiliation{QUANTUM, Institut f\"ur Physik, Universit\"at Mainz, Staudingerweg 7, 55128 Mainz, Germany}

\author{Thomas Kornher}
\affiliation{3. Physikalisches Institut, Universit\"at Stuttgart, 70569 Stuttgart, Germany}

\author{Georg Jacob}
\altaffiliation{Present address: Alpine Quantum Technologies GmbH, c/o Greiter Pegger Kofler \& Partner, Maria-Theresien-Straße 24, 6020 Innsbruck, Austria}
\affiliation{QUANTUM, Institut f\"ur Physik, Universit\"at Mainz, Staudingerweg 7, 55128 Mainz, Germany}

\author{Felix Stopp}
\affiliation{QUANTUM, Institut f\"ur Physik, Universit\"at Mainz, Staudingerweg 7, 55128 Mainz, Germany}

\author{Samuel T. Dawkins}
\affiliation{Experimentalphysik I,
Institut f\"ur Physik, Universit\"at Kassel, Heinrich-Plett-Stra{\ss}e 40, 34132 Kassel, Germany}

\author{Roman Kolesov}
\affiliation{3. Physikalisches Institut, Universit\"at Stuttgart, 70569 Stuttgart, Germany}

\author{J\"org Wrachtrup}
\affiliation{3. Physikalisches Institut, Universit\"at Stuttgart, 70569 Stuttgart, Germany}

\author{Kilian Singer}
\affiliation{Experimentalphysik I,
Institut f\"ur Physik, Universit\"at Kassel, Heinrich-Plett-Stra{\ss}e 40, 34132 Kassel, Germany}

\author{Ferdinand Schmidt-Kaler}
\affiliation{QUANTUM, Institut f\"ur Physik, Universit\"at Mainz, Staudingerweg 7, 55128 Mainz, Germany}

\begin{abstract}
Single dopant atoms or dopant-related defect centers in a solid state matrix provide an attractive platform for quantum simulation of topological states~\cite{perczel2017topological}, for quantum computing and communication, due to their potential to realize a scalable architecture compatible with electronic and photonic integrated circuits~\cite{dibos2018atomic,zhong2018optically,kornher2017amorphous,marzban2015observation,englund2010deterministic,lemonde2018}.
The production of such quantum devices calls for deterministic single atom doping techniques because conventional stochastic doping techniques are cannot deliver appropriate architectures. Here, we present the fabrication of arrays of praseodymium color centers in YAG substrates, using a deterministic source of single 
laser-cooled Pr$^+$ ions. The beam of single Pr$^+$ ions is extracted from a Paul trap
and focused down to 30(9)\,nm. Using a confocal microscope 
we determine a conversion yield into active color centers up to 50\% and realizing a placement accuracy of better than 50\,nm.
\end{abstract}

\maketitle
Deterministic doping methods at the nm-scale provide a route towards scalable quantum information
processing in solid state systems.
Prominent examples of atomic systems in solid state hosts 
for quantum computing are single phosphorus atoms in silicon~\cite{pla2012single,koch2019} and 
spin correlated pairs of such donors~\cite{veldhost2015,broome2018} which have led to studies of the scalability of large arrays of coupled donors~\cite{pla2012single}. Alternatively, single color
centers~\cite{Gruber2012} and the growing variety of single rare-earth ions (REI) doped into
crystalline hosts have also been employed~\cite{kolesov2012optical,kolesov2013mapping,yin2013optical,
utikal2014spectroscopic,dibos2018atomic,zhong2018optically}. Driven by proposed quantum applications, the need to
deterministically place single dopants into nanostructured devices has led to the development 
of various techniques related to the silicon material system~\cite{van2015single,ruess2007}.
Crystalline hosts of color centers and REI, however, typically exhibit poor electronic properties, which inhibits single ion detection via active substrates~\cite{van2015single} and therefore an alternative technique for deterministic implantation of dopants is required.
Here, we present an inherently deterministic method for single ion implantation based on a 
segmented Paul trap which allows for implantation in any solid state material with a broad range of implantation energies.

For characterizing the implantation method, we use single praseodymium ion detection in yttrium aluminum garnet (YAG) crystals based on upconversion microscopy. This detection scheme requires implanted
praseodymium ions to arrange in the proper lattice position and reach the Pr$^{3+}$ charge state through a suitable annealing and
activation procedure. An accurate determination of the ratio of detected ions to 
implanted ions, commonly referred to as implantation yield, has been performed for the first 
time at the level of single ions and will further foster the optimization of annealing procedures. 
In comparison to previous implantation-based nitrogen and silicon vacancy color center generation experiments~\cite{schroeder2017}, we achieve more than 20 times higher yield for the implantation of Pr$^+$ in YAG, even at much lower implantation energies with correspondingly smaller straggling-related uncertainty of the implantation site.
The letter is organized as follows: After introducing the apparatus and procedures for 
deterministic implantation, we characterize the Pr$^{3+}$ centers in YAG samples through confocal 
two-photon microscopy imaging. We discuss the spatial uncertainty of both single, and arrays 
of REI-generated color centers, also sketching further plans for applications and improvements. 

\begin{figure*}[htb]
\begin{center}
\includegraphics[width=0.8\textwidth]{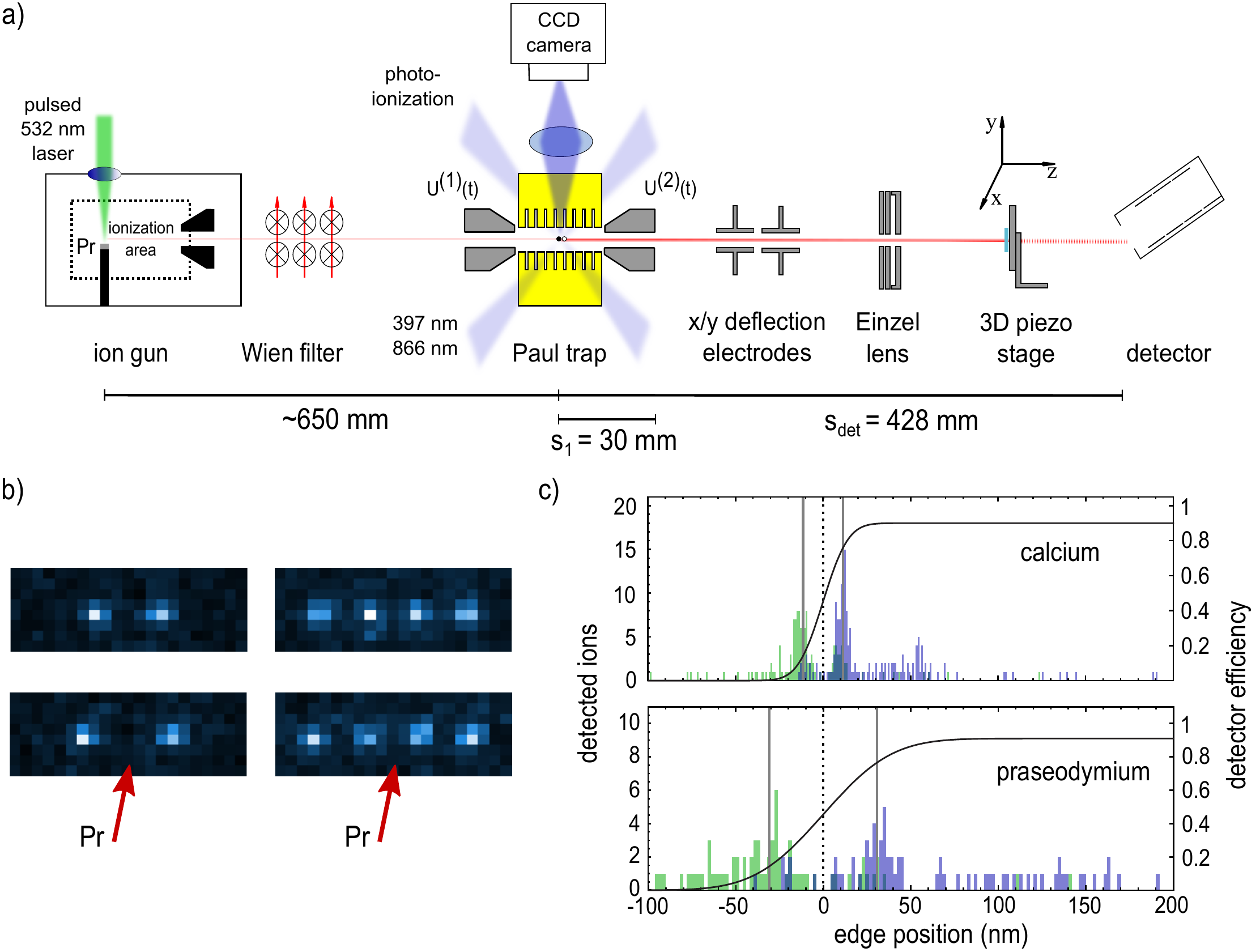}
\caption{a) Sketch of the single ion implantation setup. b) Fluorescence of ions imaged. Top: Pure Ca$^+$ crystals, the distance between two Ca$^+$ ions is 9.5\,$\mu$m. Bottom: crystals containing an additional Pr$^+$ ion. c) Histograms of the profiling edge measurement for $^{40}$Ca$^+$ and $^{141}$Pr$^+$ ions. For calcium, the extraction took about 15\,min for 308 ions, whereas for praseodymium it took about 2\,h for 150 ions. The events of the single ion extraction are split into two cases, blue presents the detected ions and green the blocked ions. The black line shows the Bayesian fit function which corresponds to the last measured parameter values for beam position $x_0$, radius $\sigma$ and detector efficiency $a$. The mean value of the beam position of the x-axis is set to $x_0=0$ (dotted line) and the gray lines show the 1-$\sigma$ radius of the beam waist, which is $\sigma_{\text{Ca}} = 11.3 \pm 2.0$\,nm for calcium and $\sigma_{\text{Pr}} = 30.7 \pm 8.5$\,nm for praseodymium.
\label{fig:img1}}
\end{center}
\end{figure*}

At the heart of the experimental apparatus is an ion trap, which acts as a source of single $^{141}$Pr$^+$ ions. The source is realized by loading the ions into the trap, where they are cooled, identified and subsequently extracted towards the implantation section (see Fig.~\ref{fig:img1}a).
The ion trap consists of two segmented DC and RF-electrodes in an X-shaped configuration 
with two endcap electrodes at a distance of 2.9\,mm~\cite{jacob2016transmission}. We operate the trap
with a radio frequency of $\Omega_{\text{RF}}=2\pi\times 23.062$\,MHz at a peak-to-peak amplitude
of V$_{\text{pp}}=572$\,V, which leads
to $\omega_{\text{ax,r1,r2}}=2\pi\times\{$0.45,\,1.584,\,1.778$\}$\,MHz for the axial and
radial mode frequencies of a single $^{40}$Ca$^+$ ion. We load mixed crystals
of $^{141}$Pr$^+$ and  $^{40}$Ca$^+$ ions, and employ laser cooling on the
S$_{1/2}$ to P$_{1/2}$ transition of Ca$^+$ to sympathetically cool the Pr$^+$ ions (for details see Methods M1). By imaging the fluorescence of the calcium ions on a camera we determine the number of trapped ions.
The increase of distance between two bright calcium ions provides evidence for a single praseodymium ion trapped between them (see Fig.~\ref{fig:img1}b). 
We reduce the ion number to exactly one $^{40}$Ca$^+$ and one $^{141}$Pr$^+$ by a predefined voltage sequence 
of the axial trap potential. 
The ions are extracted with an energy of 5.9\,keV. The extraction path is steered by deflection electrodes to the center of an electrostatic einzel lens which focuses the ions to a small spot. 
A piezo translation stage, which can be moved in the focal plane is used to determine the spot size with a profiling edge and allows for precise positioning of praseodymium implantation with respect to the YAG crystal. The profiling edge measurement for calcium and praseodymium is shown in Fig.~\ref{fig:img1}c.
The measured radius of the beam waist for calcium is $\sigma_{\text{Ca}} = 11.3\pm 2.0$\,nm 
and $\sigma_{\text{Pr}} = 30.7\pm 8.5$\,nm in the case of praseodymium (for details see M2 - M4). For
the case of Doppler-cooled Ca$^+$ ions with a measured wavepacket size of about 52\,nm, 
we found that the spot size is dominated by mechanical 
vibrations~\cite{jacob2016transmission}. To understand the spot measurements for
Pr$^+$, however, we conjecture an elevated phonon number of radial modes, corresponding to an increased motional wavepacket size, because the sympathetic cooling rate is
significantly reduced for differences in the mass-to-charge ratio as high as 141/40 
$\simeq$ 3.52~\cite{wuebbena2012}. A YAG crystal was 
placed in the focal plane and implanted with two dot-grid patterns of praseodymium ions, each with 12 
spots with exactly eight (area A) and exactly four (area B) Pr$^+$ ions per spot,
respectively. The dot-grid spacing was 2\,$\mu$m and the implantation energy 
of 5.9\,keV corresponds to an implantation depth of $\sim$\,6\,nm \cite{srim}. 
After implantation the sample was annealed in air at 1200\,$^\circ$C for approximately 1\,min.
\par
\begin{figure*}[htb]
\centering
\includegraphics[width=0.80\textwidth]{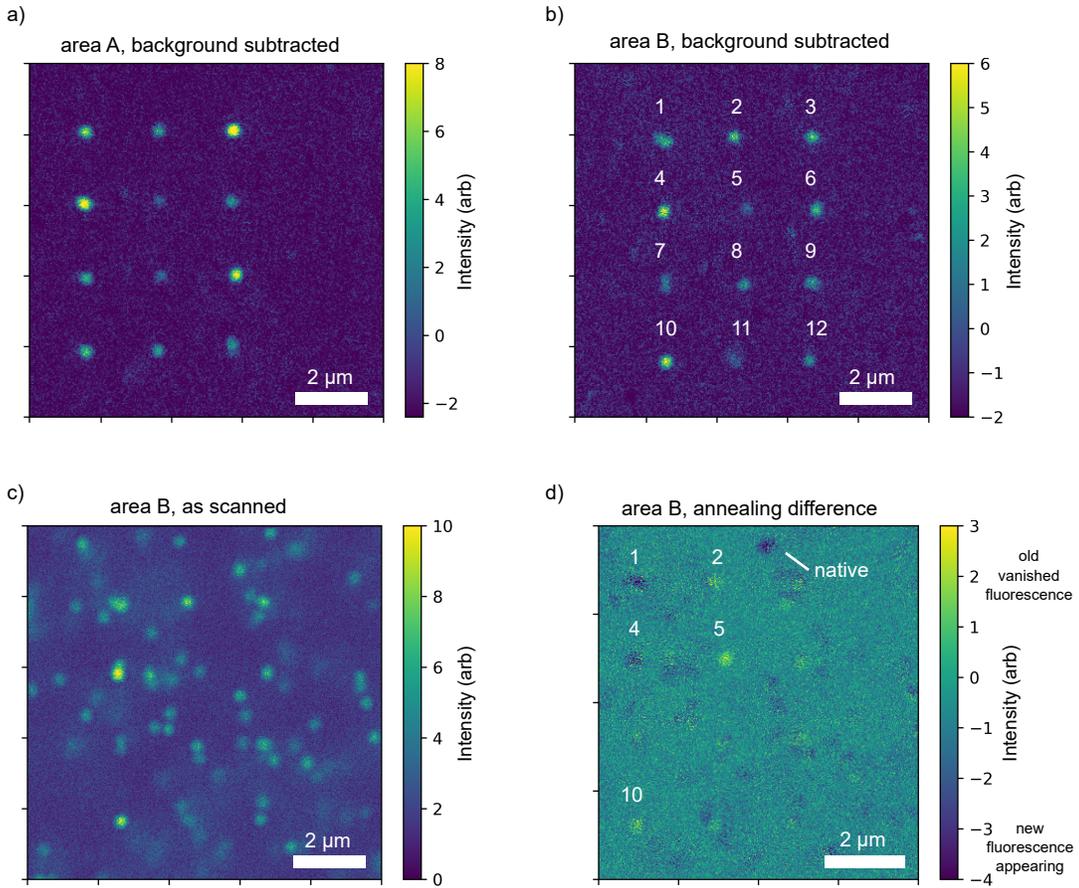}
\caption{Upconverting microscopy scans of a) Area A, after subtraction of pre-implantation
scanned background. b) Area B, after subtraction of pre-implantation scanned background.  
c) Area B, after implantation and annealing procedure. d) Difference in REI fluorescence signals $\Delta C = C_{\scriptsize\mbox{before}} - C_{\scriptsize\mbox{after}} $, before and after a second annealing process.\label{fig:img3}}
\end{figure*}

Verification of successful color center generation was demonstrated by optical detection of trivalent praseodymium ions on the single ion 
level~\cite{kolesov2012optical}. The method is based on a two-photon
upconversion~\cite{gayen1992two}, which enhances the efficiency of the excitation-emission 
cycle and maximizes the fluorescence emission. The electronic level structure of Pr$^{3+}$ 
ion in the YAG crystal~\cite{gruber1989symmetry} allows for several two-photon excitation 
schemes (see Fig.~\ref{fig:img2}a).
\begin{figure}[tb]
\centering
\includegraphics[width=1\columnwidth]{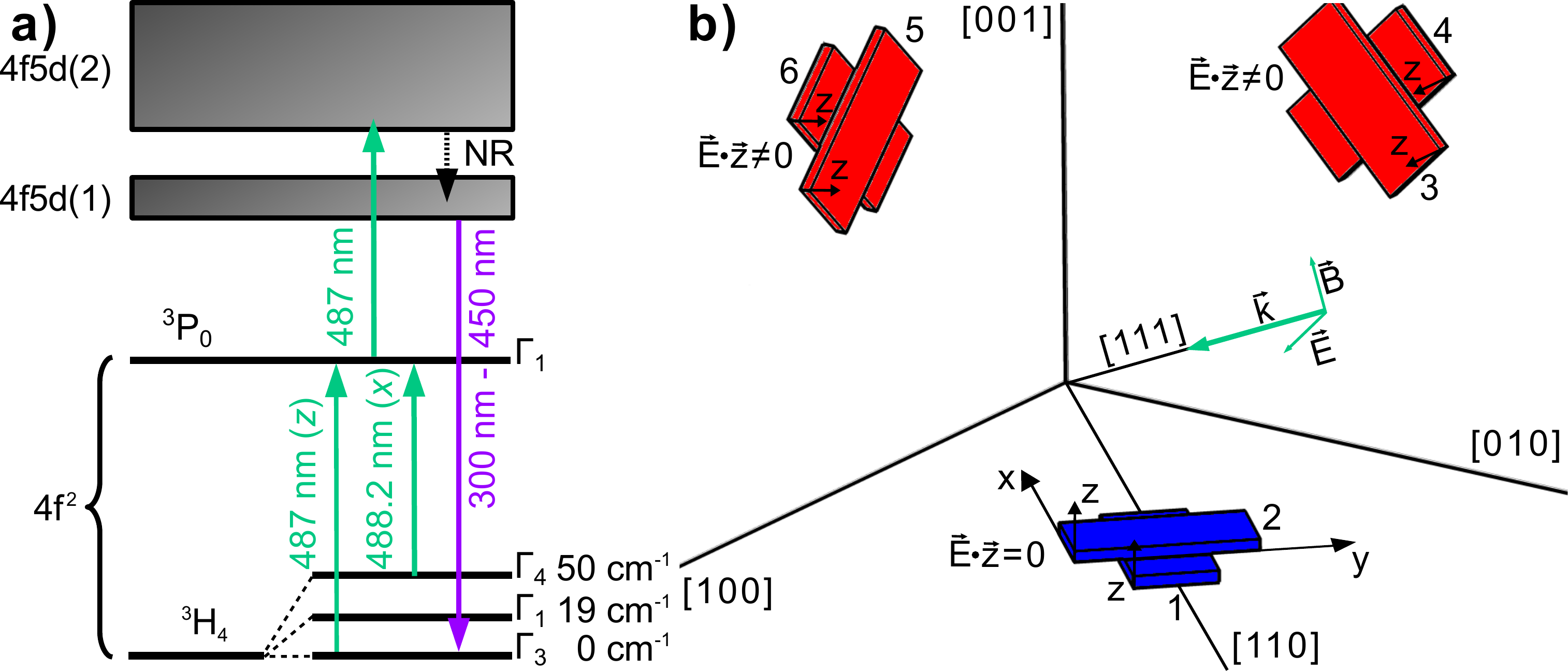}
\caption{a) Electronic level structure of Pr$^{3+}$ in a YAG crystal. For optical
characterization, the blue transition (487\,nm) originating from $\Gamma_3$ and 
exciting the $z$ dipole directions was used. Levels have symmetry representations 
$\Gamma_3$ (0\,cm$^{-1}$), $\Gamma_1$ (19\,cm$^{-1}$) and  $\Gamma_4$ (50\,cm$^{-1}$).
Transitions from $^{3}$H$_4$  to $^{3}$P$_0$ state ( $\Gamma_1$ symmetry) are polarized
either along the $z$ axis or along the $x$ axis. b) Orientation of the six dodecahedral
rare-earth sites, indicated by matchboxes~\cite{dillon1961jf}. The excitation 
laser propagates along the [111] and its polarization is adjusted so that two out of 
six sites are rendered dark. As an example, sites 1 and 2 are not excited, which leads
to sites 3 to 6 being excited with equal, nonzero probability. 
\label{fig:img2}}
\end{figure}
We employ a laser near 487\,nm driving a parity-forbidden, 
thus spectrally narrow,  $4f-4f$ transition from $^{3}$H$_4$ ground state to $^{3}$P$_0$ 
shelving state with a lifetime of  8\,$\mu$s. From $^{3}$P$_0$ state, a second excitation 
step at 487\,nm promotes the electron to the $4f5d$ level via a parity-allowed optical 
transition. The lowest $4f5d$ level yields emission in a 
spectral range between $300-450$\,nm\cite{gayen1992two}, with close to unity quantum 
efficiency \cite{weber1973nonradiative} and a lifetime of about 18\,ns. Such two-photon
upconversion microscopy has the advantage of virtually background-free imaging, and was 
realized by an upconversion microscope shown schematically in the supplementary material \cite{supplement}. Its optical resolution was determined to be 115(3)\,nm, from the average width of a 2D-Gaussian fit of single Pr$^{3+}$ ion fluorescence. The imaging quality is limited by background fluorescence from other Pr$^{3+}$ impurities, naturally found in the crystal within $1-2$ microns below the surface, as shown in Fig.~\ref{fig:img3}a. 
We determine a background density of Pr$^{3+}$ ions of  
6$\times$10$^{11}$\,cm$^{-3}$ or 0.04\,ppb relative to yttrium.
Pr$^{3+}$ ions doped into YAG substitute for Y$^{3+}$. The 
crystal features six magnetically inequivalent orientations of these particular sites of D$_2$ 
symmetry, where
the local $x$, $y$ and $z$ axes of the Pr$^{3+}$ ion correspond to the [110], 
[1$\overline{1}$0] and [001] crystal axes and their six equivalent directions 
(see Fig.~\ref{fig:img2}b)~\cite{van1971optical}. The $^{3}$H$_4$ ground state of
Pr$^{3+}$ in crystal field is split into nine levels, of which the lowest three in energy are populated at room temperature. 

In order to quantify single ion implantation, we render all detected Pr$^{3+}$ ions equal
with respect to the collected fluorescence signal. The linear polarization of the
excitation light near 487\,nm is switched to all three different polarization orientations,
such that two out of six Pr$^{3+}$ sites appear dark, while four states yield equal fluorescence,
see Fig.~\ref{fig:img2}b. This allows for extracting the number of ions residing within
the optically resolved spot, for details see supplementary material~\cite{supplement}.
Fig.~\ref{fig:img3}c shows an image of fluorescence of area B, comprising twelve spots each implanted with four praseodymium ions (spot no. 12 of area B has two implanted ions). However, here the implantation pattern is neither clearly visible, nor allows for a quantitative analysis on this scanned map, because implanted and pre-existing native Pr$^{3+}$ are indistinguishable. Therefore, a background
fluorescence image was scanned prior to the Pr$^+$ ion implantation, in a designated 
field of implantation. After implantation, annealing and imaging, a background subtraction
shows the newly generated Pr$^{3+}$ sites, see Fig.~\ref{fig:img3}b. The same 
procedure was used to image generated REI sites in a second YAG area A, see Fig~\ref{fig:img3}a. 

In order to study the effect of the annealing procedure on the REI fluorescence of both native
and implanted Pr$^{3+}$ ions, the implanted areas were imaged again after a second annealing step. The  difference image is shown in Fig.~\ref{fig:img3}d, taken on area B. 
Changes due to the second annealing process are visible, typically at the implantation sites and
are marked by a number. In spots two, five and ten, respectively, the fluorescence of 
a single Pr$^{3+}$ ion vanished, while in spots one and four, exactly one additional 
Pr$^{3+}$ ion appeared. Only one single native Pr$^{3+}$ ion appeared on a 100\,$\mu$m$^2$
area, leading to the conclusion, that the annealing procedure has only a marginal effect 
on the native Pr$^{3+}$ background and our background correction is a valid procedure.
We conjecture that the diffusion rate of 2$\times$10$^{-21}$\,m$^{2}$/s at a temperature 
$\sim$1200$^{\circ}$C is responsible for this effect~\cite{cherniak1998rare}. For shallow 
locations, 6\,nm below the surface, Pr$^{3+}$ ions may diffuse and stabilize at the 
surface in a non-fluorescing charge state. Shorter annealing times may help to reduce this effect
but were not available for our experiments.
The enhanced mobility of Pr$^{3+}$ ions in implanted spots as compared to native Pr$^{3+}$ is likely caused by local crystal damage in these sites that occurs even at low implantation energy of 5.9\,keV and single 
ion fluence.
\par
\begin{figure}[htb]
\centering
\includegraphics[width=1\columnwidth]{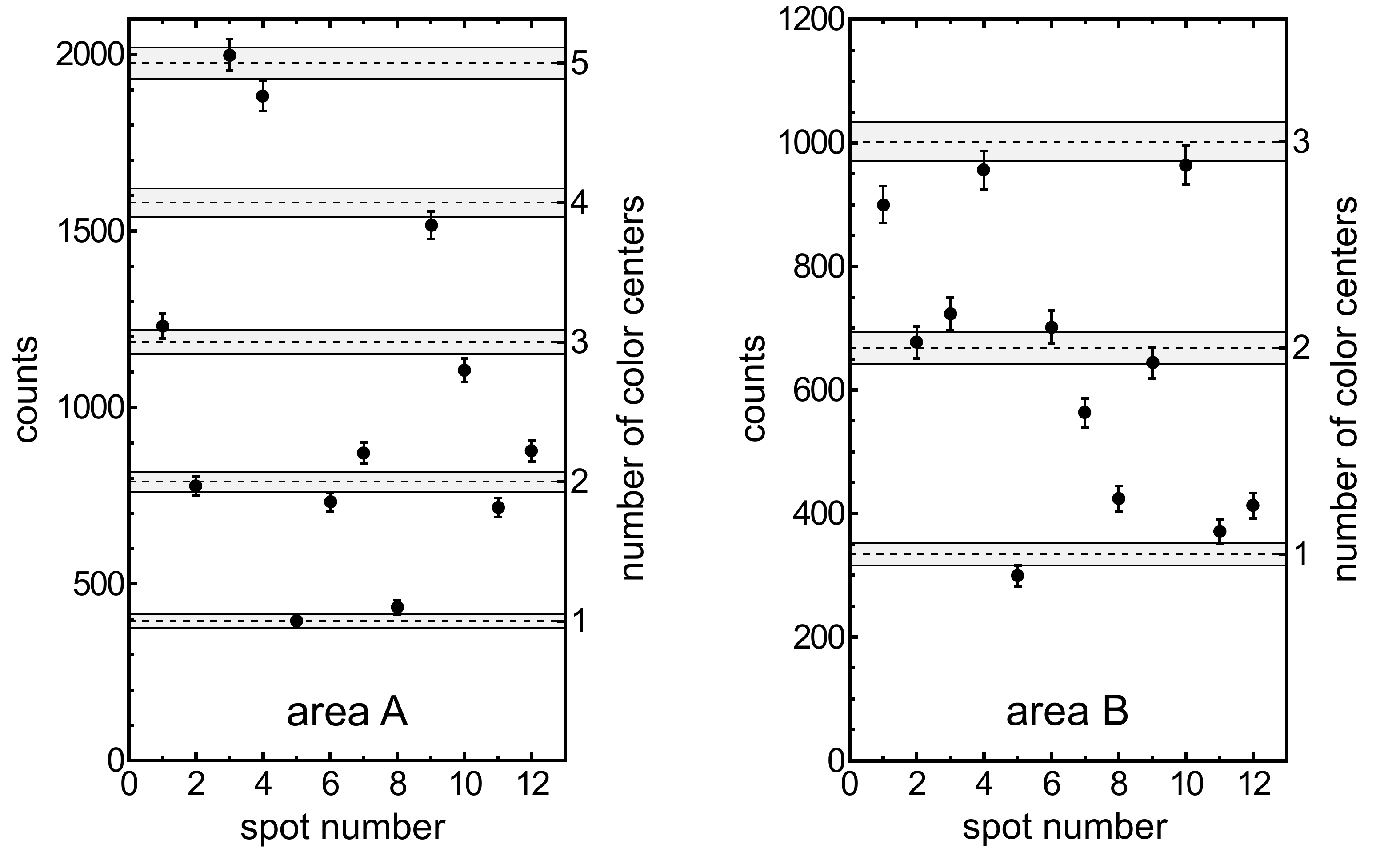}
\caption{Net fluorescence signal (per 6\,ms and after subtraction of background) of implanted spots for the areas A and B. The observed count-rate is consistent with the discrete nature of the integer number of REI emitters per spot. For the error bars we assume a Poissonian photon counting statistics.
\label{fig:img4}}
\end{figure}

Summing up the collected REI fluorescence per spot, after background subtraction, we find approximately integer multiples of single Pr$^{3+}$ ion counts, see Fig.~\ref{fig:img4}.
Accordingly, each spot is assigned an integer number of 
optically active Pr$^{3+}$ ions, see supplementary material \cite{supplement}. The 
associated implantation yields are  32\,\% and 50(5)\,\% for area A and area B,
respectively.
We account only for systematic errors, estimated from the observed annealing-induced ion migration in area B.
We conjecture that the lower yield in area A may be attributed to the twice higher implantation dose as compared to area B, thus a higher probability of lattice defects which have not been fully annealed. 

\begin{figure*}[htb]
\centering
\includegraphics[width=0.75\textwidth]{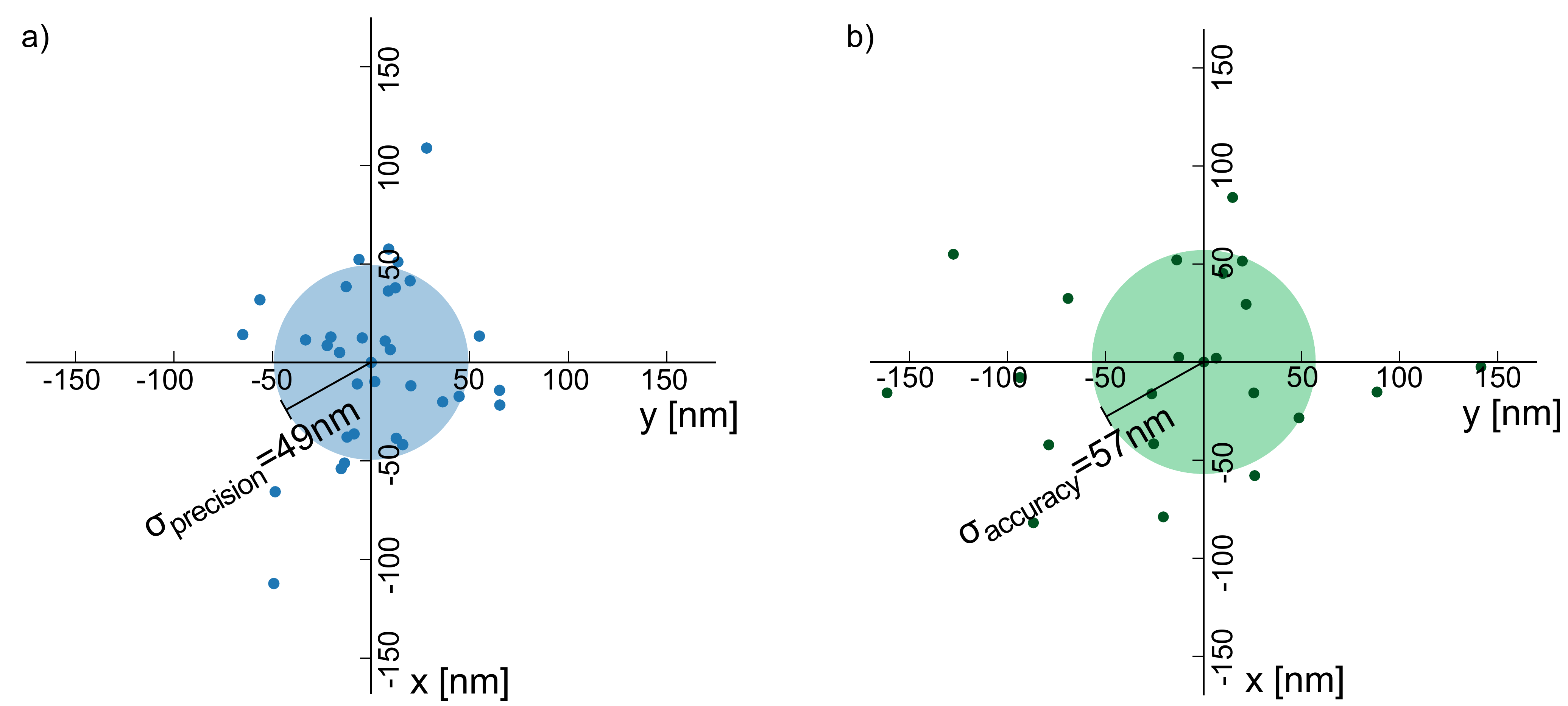}
\caption{a) Fitted Pr$^{3+}$ ion locations with respect to the centroid of each respective 
spot. For this analysis we excluded the implantation spots
1, 7 and 9 in area B and spot 12 in area A, since these suffer from drifts far 
larger than the standard deviation of the rest of the implanted spots. 
The standard deviation of $\sigma_{\text{precision}}=49$\,nm is an indication of the precision of ion
placement.
b) Calculated difference in position between the implanted spots and the ideal grid
positions. 
The standard deviation of $\sigma_{\text{accuracy}} = 57$\,nm is an indication
of the accuracy of ion placement. 
\label{fig:img5}}
\end{figure*}

We study the spatial quality of placing REI in the spots: Fitting 2D-Gaussian profiles onto
native single Pr$^{3+}$ fluorescence images reveals an optical circular single ion point
spread function (PSF) of the confocal microscope of 115\,nm. Setting this width, we fit 
each of the implanted spot with a multi-Gaussian, according to the number of REI in that 
spot. The extracted positions of fitted PSFs are displayed relative to the center of mass of the respective spot in Fig.~\ref{fig:img5}a. 
The variance of all PSFs is $\sigma_{\text{precision}}=49$\,nm 
and indicates a measure of the position-stability, as a scatter around the true value. Note, that the Pr$^+$ ion beam size 
of about 30\,nm dominates the position-stability, while straggling uncertainties and 
annealing-induced migration further add to this error. The deviation of implantation spots off
the desired grid is characterized as a patterning-accuracy. For this, we extract the 
center of each fluorescent spot from a 2D-Gaussian fit (independent of the number of color 
centers in the spot) and calculate its difference to an ideal grid position.
The standard deviation of $\sigma_{\text{accuracy}} = 57$\,nm characterizes the accuracy for
array-writing, see Fig.~\ref{fig:img5}b. We conjecture that this accuracy is additionally
affected by long-term thermal drifts and piezo-actuator inaccuracies. 

In conclusion, we have demonstrated deterministic ion implantation and the writing of an
array of Pr$^{3+}$ ions in YAG. The characterization of 
implantation sites by two-photon confocal microscopy shows a yield of Pr$^{3+}$ ions 
in YAG of up to 50\%. Improving the annealing procedures could allow for approaching $100\%$ and thus deterministic generation of REI dopants in crystals. 
Furthermore, we could demonstrate a 50\,nm positioning-accuracy, entering the range which is required
for efficient coupling to photonic structures or electrical antennas to couple single spins with
supeconducting circuits. 
Improving sympathetic cooling of the Pr$^+$ ions prior to extraction, as well as increasing the mechanical and thermal stability of the apparatus, we anticipate improving the positioning-accuracy to a few nanometers, a regime which is dominated by the implantation induced straggling.
Using superresolution microscopy, which is 
available in the case of Ce$^{3+}$ ions in YAG~\cite{kolesov2018superresolution}, a more
precise characterization of the implantation systems can be implemented. Note, that the 
single ion implantation technique may be applied for a large range of materials, doping 
ions and implantation energies, which gives rise to new  options for fabricating quantum devices, such as arrays of phosphorus qubits in an ultra-pure silicon crystal.
 
\subsection{Methods and Materials}
\paragraph{M1: Generation and loading of Pr$^+$ ions into the Paul trap:}
For generating $^{141}$Pr$^+$ ions a commercial ion gun (Specs IQE 12/38) is modified with
a ceramic stick holding a praseodymium target (1mm thick, purity 99.5\%) inserted into the repeller electrode structure  (Fig.~\ref{fig:img1}a, black dotted). A single pulse from a Nd:YAG laser (532\,nm, 1.2\,mJ), ablates praseodymium which is ionized by electron impact. Ions are accelerated to 600\,eV by an extraction electrode (Fig.~\ref{fig:img1}a, black) and steered with a Wien filter through the pierced endcap (hole \(\varnothing\)=200\,$\mu$m, length 28.55\,mm). To decelerate the ions, a repulsive potential of $\text{U}^{(1)}(t)=350$\,V is applied to this first endcap for $40.9\,\mu$s. Inside the endcap bore, the ions are shielded from the electrical fields. At $t=41\,\mu$s, the endcap is switched to $\text{U}^{(1)}(t)=-256$\,V, decelerating the ions when they leave the endcap bore. Simultaneously, the second trap endcap is set to a repulsion potential of $\text{U}^{(2)}(t)=+3$\,kV to prevent the ions from leaving the trap volume. After 70\,ns both endcaps are switched to ground potential, the DC-segments are used to form a confining axial potential at $\omega_{\text{ax}}=450.8$\,kHz. We achieve an average rate of 1 ion per min for loading, identification, cooling and extracting single $^{141}$Pr$^+$ ions. Calcium and praseodymium ions are loaded simultaneously: Random numbers of calcium ions are loaded by photoionization from a resistively heated oven, not shown in Fig.~\ref{fig:img1}a, and laser cooled on the S$_{1/2}$ to P$_{1/2}$ transition, while praseodymium ions are injected.

\paragraph{M2: Extraction of Pr$^+$ ions:}
The ions are extracted towards the sample by applying an acceleration voltage of $\text{U}^{(2)}(t)=-5.9$\,kV to the second endcap. Near this endcap, the RF field would have strong axial components and would alter the ion's kinetic energy, causing chromatic aberrations of the spot. In order to mitigate this, the extraction process is phase-synchronized with the RF drive, switching the RF off at the instance of extraction. For this we 
%$\pi$-phase-shift 
invert the RF at an amplitude higher by +6\,dB to foster destructive interference in the resonance circuit supplying the trap electrode. Furthermore, we switch off the RF input at the zero crossing point. We jointly extract calcium and praseodymium ions. Due to their different $m/q$-ratio, they have different trajectories and the Ca$^+$ ions hit at the target plane a few $\mu$m away from the Pr$^+$ implantation site.

\paragraph{M3: Spot size measurement:} A cleaved GaAs crystal knife edge mounted on the 3D-piezo translation stage is moved into the ion beam, while the ion transmission signal is recorded by a secondary electron multiplier (SEM) detector, see Fig.~\ref{fig:img1}a. We use the Bayesian experimental design method~\cite{jacob2016transmission}, an algorithm that maximizes the information gain by calculating the ideal knife edge position for each ion extraction event. From the time-of-flight signal we verify the detection of Pr$^+$ ions, see supplementary material~\cite{supplement}. Such destructive detection method confirms the \textit{in situ} Pr$^+$ identification.

\paragraph{M4: Non-destructive praseodymium ion identification}
Imaging the fluorescence of the $^{40}$Ca$^+$ ion crystal by means of an electron multiplying charge-coupled device (EMCCD) camera allows for identification of Pr$^+$ ions. In the case of $n=2$ calcium ions, the distance between calcium ions is increased, see Fig.~\ref{fig:img1}b, lower row, as compared to a calcium crystal, upper row. This shift is harder to observe for $n\geq 4$ ions, so in this case we reduce the number of calcium ions by a voltage sequence on the DC segments in order to ensure unambiguous identification of praseodymium.

\subsubsection{Selection rules of Pr$^{3+}$:YAG}
The local $z$ axis conventionally coincides with [001]. Based on the D$_2$ point group symmetry 
operations, under which the Pr$^{3+}$ wavefunctions must be invariant, D$_2$ has four 
irreducible representations $\Gamma_1$, $\Gamma_2$, $\Gamma_3$ and $\Gamma_4$. 
Electric-dipole
transitions between states with different representation follow the selection 
rules listed in Table \ref{tab:table1}
and are polarized along the indicated site 
axis\cite{gruber1989symmetry}. 
\setlength{\tabcolsep}{5mm}
\renewcommand{\arraystretch}{1.2}
\begin{table}[h!]
  \centering
  \caption{Electric-dipole selection rules for D$_2$ symmetry}
	\vspace{0.2cm}
  \label{tab:table1}
  \begin{tabular}{|c|c|c|c|c|}

  \hline
  &   $\Gamma_1$  & $\Gamma_2$ & $\Gamma_3$ & $\Gamma_4$\\\hline
$\Gamma_1$ &   - &  y& z & x\\    \hline 
    
$\Gamma_2$  &  y & - & x&z\\\hline
$\Gamma_3$ &  z& x & -&y\\\hline

$\Gamma_4$& x  & z & y &-\\\hline

  \end{tabular}
\end{table}
The absorption cross-section for an individual dipole is given by the
projection of its direction onto the electric field vector of the excitation
light. For our [111] cut YAG crystal the electric field vector is oriented
in the plane perpendicular to the [111] direction and circularly polarized 
excitation light can theoretically excite all six magnetically inequivalent
sites along the $z$ dipole with equal efficiency. For half of the $x$ dipoles the 
excitation efficiency is always three times less than for the other half and
therefore the $x$ polarized transition cannot be used for quantification. 
In order to avoid systematic errors induced by a non-ideal circular
polarization, for example due to potential distortion by the objective lens or dichroic filter,
linearly polarized excitation light was used. Since two out of six $z$ dipoles 
are always aligned parallel, we are left with a threefold symmetry for 
[111] cut YAG crystals with respect to linear polarized excitation light. 
Accordingly, we can render two out of six sites dark, while the other
four fluoresce with the same brightness.
Collecting fluorescence data for three distinct excitation polarizations, such 
that two out of six sites are always dark, then reliably yields the 
same fluorescence signal for all of the six possible sites when added 
up. With this method, all Pr$^{3+}$ ions residing in the same plane yield
comparable fluorescence signal, needed for quantification. 
Optical resolution was determined to be $c=115(3)\,$nm, from the average width of a two 
dimensional Gaussian fit onto single Pr$^{3+}$ ions of the form:

\begin{equation*} 
   \begin{split} 
f(x,y) = &A \cdot \exp \Bigg(\frac{(\cos(\beta)\cdot (y-y_0)+\sin(\beta)\cdot (x-x_0))^2}{2c^2} \\
&- \frac{(\sin(\beta)\cdot (y-y_0)+\cos(\beta)\cdot (x-x_0))^2}{2c^2} \Bigg)+B
   \end{split} 
\end{equation*} 

\subsubsection{Upconversion microscopy of Pr$^{3+}$:YAG}
Two-photon upconversion microscopy has the advantage of virtually
background-free imaging, because the detected wavelength can be well seperated from 
the strong flux of excitation photons and especially from background fluorescence
originating from fluorescing impurities other than Pr$^{3+}$.
The typically required pinhole for high 
resolution microscopy in confocal setups can be omitted in the deployed microscope
setup, shown in figure \ref{fig:img2}, since the non-linear 
intensity 
dependence of the upconversion mechanism confines the focal plane comparably well. In Table
\ref{tab:table2} and \ref{tab:table3}, we show the fluorescence signal on the measured
spots in area A and area B, showing steps of multiples of the single Pr$^{3+}$ count rate.
\begin{figure}[htb]
\includegraphics[width=1\columnwidth]{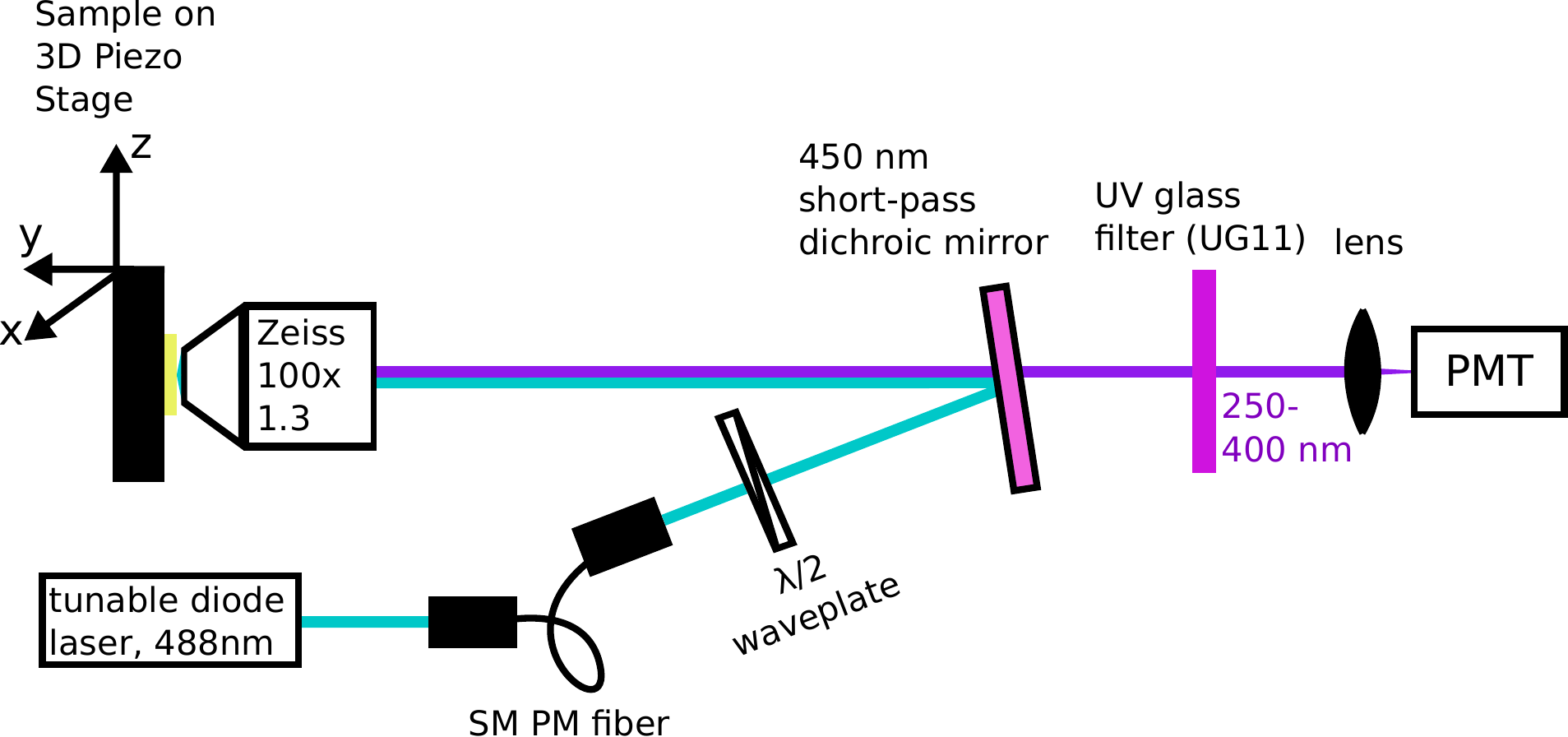}
\caption{Experimental upconverting microscope setup for detection of single Pr$^{3+}$ in
YAG. PMT: Photomultiplier Tube (Hamamatsu H11890-210, 30\% detection efficieny at 400\,nm). 
The 450\,nm short-pass dichroic mirror had at least OD4 and the UG11 filter 
(250-400\,nm bandpass) had at least
OD5 for wavelengths longer than 410\,nm. The 488\,nm tunable diode laser is comprised of a 
home-built laser diode power supply and tunability is achieved through prism feedback, as 
described in \cite{kolesov2012optical}. The laser is coupled into a single-mode (SM) 
polarization maintaining (PM) fiber. 
\label{fig:img2}}
\end{figure}

\subsubsection{Annealing procedure of YAG crystals}
After implantation of the sample, the following annealing procedure was conducted in air atmosphere:\\
Ramping up to 800$^{\circ}$C at rate of 5$^{\circ}$C/min. Then ramping up to 1200$^{\circ}$C 
at rate of 3$^{\circ}$C/min. Hold 1 minute at 1200$^{\circ}$C. Then ramping down to 
800$^{\circ}$C at rate of 3$^{\circ}$C/min. Afterwards cooling down to room temperature at
uncontrolled rate.

\setlength{\tabcolsep}{5mm}
\renewcommand{\arraystretch}{1.2}
\begin{table}[htb]
\small
  \centering
  \caption{Accumulated Fluorescence signal on spots in area A. Sums take into account a circular cropped out region of
  pixels of the raw collected data on the respective regions minus the average background in 
  proximity of this crop out defined by a donut shaped outer lying 
  second crop out of raw pixels. Pixel size is 25\,nm and pixel integration time is 6\,ms 
  on all taken scans.  Average summed up single ion fluorescence counts are 
  $C_A=395$ kcounts and corresponds
  to the area under the fitted 2D-Gaussian.
 }
\vspace{0.2cm}
  \label{tab:table2}
  %\begin{tabular}{|c|c|c|c|}
\begin{tabular}{|p{0.8cm}|p{1.2cm}|p{1.2cm}|p{1.2cm}|}
  \hline
 Spot no. &   Spot count  & Ions per spot & Rounded \\
  & [kcounts]& [in units of $C_A$]& [ions/spot]\\ \hline
1 &   1231 &  3.12& 3 \\    \hline    
2  &  778 & 1.97 &2\\\hline
3 &  1998& 5.06 & 5\\\hline
4& 1883  & 4.77 & 5\\\hline
5& 396  & 1 & 1 \\\hline
6& 732  & 1.85 & 2 \\\hline
7& 871  & 2.21 & 2 \\\hline
8& 434  & 1.1 & 1 \\\hline
9& 1516  & 3.84 & 4 \\\hline
10& 1105  & 2.8 & 3 \\\hline
11& 717  & 1.82 & 2 \\\hline
12& 876  & 2.22 & 2 \\\hline
Sum& -  & - & 32 \\\hline

  \end{tabular}
\end{table}

\setlength{\tabcolsep}{5mm}
\renewcommand{\arraystretch}{1.2}

\begin{table}[htb]
\small
  \centering
  \caption{Fluorescence signal on spots in area A. Sums take into account a circular crop out of
  pixels of the raw collected data on the respective regions minus the average background in 
  proximity of this crop out defined by a donut shaped outer lying 
  second crop out of raw pixels. Pixel size is 25\,nm and pixel integration time is 6\,ms 
  on all taken scans. 
  Average summed up single ion fluorescence counts are $C_B=334$ kcounts and corresponds
  to the area under the fitted 2D-Gaussian.}
	\vspace{0.2cm}
  \label{tab:table3}
  %\begin{tabular}{|c|c|c|c|}
	\begin{tabular}{|p{0.8cm}|p{1.2cm}|p{1.2cm}|p{1.2cm}|}
  \hline
 Spot no. &   Spot count  & Ions per spot & Rounded \\
 & [kcounts] & [in units of $C_B$] & [ions/spot]\\\hline
1 &   900 &  2.69& 3 \\\hline    
2  &  677 & 2.03 &2\\\hline
3 &  723& 2.16 & 2\\\hline
4& 956  & 2.86 & 3\\\hline
5& 299  & 0.9 & 1 \\\hline
6& 702  & 2.1 & 2 \\\hline
7& 563  & 1.69 & 2 \\\hline
8& 424  & 1.27 & 1 \\\hline
9& 644  & 1.93 & 2 \\\hline
10& 964  & 2.89 & 3 \\\hline
11& 371  & 1.11 & 1 \\\hline
12& 413  & 1.24 & 1 \\\hline
Sum& -  & - & 23 \\\hline
  \end{tabular}
\end{table}

\subsubsection{Detection of praseodymium ions}

The time-of-flight (TOF) difference between simultaneously extracted calcium and praseodymium ions is 2.6\,$\mu$s. For the spot size measurement of Pr${^+}$ ions, this time difference allows for deflection of the calcium ions during the flight or for rejecting the Ca${^+}$ TOF signal.
\begin{figure*}
\centering
\includegraphics[width=0.9\textwidth]{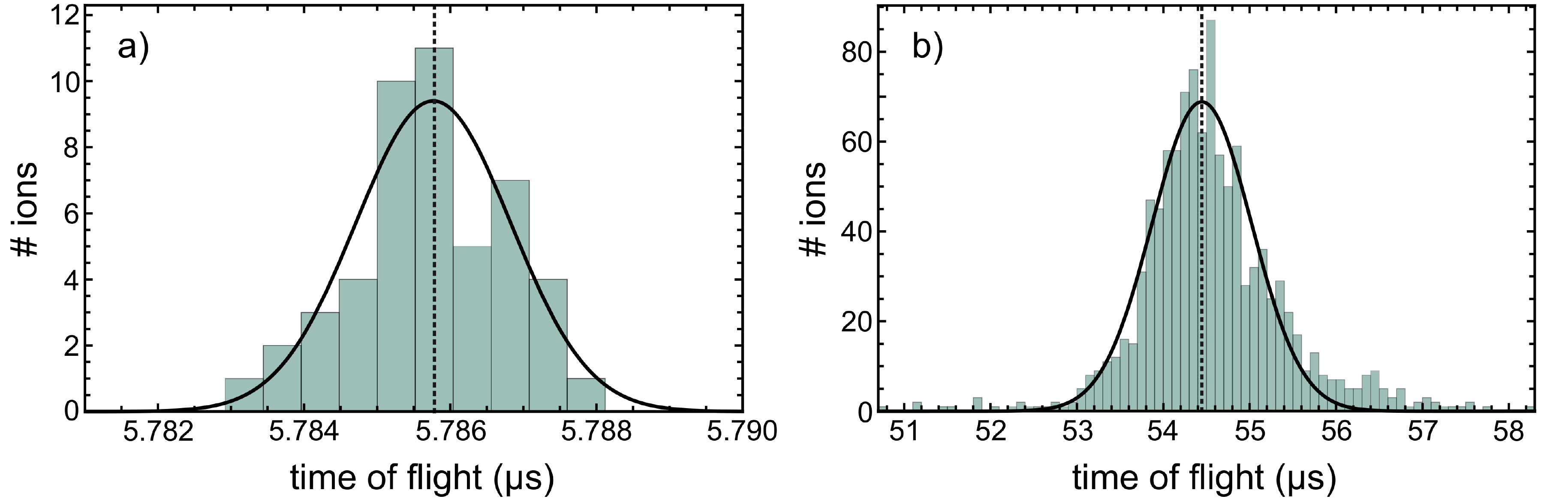}
\caption{a) TOF measurement of 50 trapped $^{141}$Pr${^+}$-ions with a FWHM spread of $\Delta t=2$\,ns and a mean velocity of 5.79\,$\mu$s. The SEM detector is $\sim$ 43\,cm away from the trap. b) Velocity distribution of laser ablated $^{141}$Pr${^+}$ ions from the ion gun, which are steered through the trap, without switching of the endcaps, onto the detector which is located at a distance of 107\,cm.
The mean velocity is 54.45\,$\mu$s with a full-width-half-maximum spread of 1.4\,$\mu$s.  
\label{fig:tof}}
\end{figure*} 
Figure \ref{fig:tof}a shows the TOF distribution for 50 trapped $^{141}$Pr${^+}$ ions with an extraction energy of 5.9\,keV. The ions have a mean velocity of 5.79\,$\mu$s with a full-width-half-maximum (FWHM) spread of $\Delta t=2$\,ns.
The SEM detector is located at a distance of 42.8\,cm from the trap. \par
Figure \ref{fig:tof}b shows the velocity distribution of laser ablated $^{141}$Pr${^+}$ ions. The Pr${^+}$ ions are steered through the trap, without switching of the endcaps for trapping, onto an SEM (secondary electron multiplier) detector which is located at a distance of 107\,cm from the ion gun. 
The mean velocity of the Pr$^+$ ions is 54.45\,$\mu$s with a FWHM spread of 1.4\,$\mu$s for an extraction energy of the ion gun of 600\,eV. In total 1085 ions were counted.

\newpage
FSK acknowledges financial support by the DFG DIP program (FO 703/2-1) and the Australian Research council within the CQC2T. FSK and KS acknowledge financial support by the VW Stiftung. RK acknowledges financial support by DFG (Grant
No. KO4999/3-1) and RK and JW acknowledge financial support by the FET-Flagship Project SQUARE, the EU via SMeL and QIA as well as the DFG via FOR
2724.

\bibliographystyle{apsrev4-1}
\bibliography{bibfile}

\end{document}